\documentclass{aa}
\usepackage{graphics}

\begin{document}      

   \thesaurus{03         
	      (11.09.1 NGC 4548; 
	       11.09.2;  
	       11.09.4;  
	       11.11.1)  
              }
   \title{Kinematics of the anemic cluster galaxy NGC 4548}

   \subtitle{Is stripping still active ?}

   \author{B.~Vollmer\inst{1,3}, V.~Cayatte\inst{3}, A.~Boselli\inst{4,5}, C.~Balkowski\inst{3} \and W.J.~Duschl\inst{1,2}}

   \offprints{B.~Vollmer, e-mail: Bernd.Vollmer@obspm.fr}

   \institute{Institut f\"ur Theoretische
              Astrophysik der Universit\"at Heidelberg, Tiergartenstra{\ss}e 15,
              D-69121 Heidelberg, Germany. \and
              Max-Planck-Institut f\"ur Radioastronomie, Auf dem H\"ugel 69,
   	      D-53121 Bonn, Germany. \and
	      Observatoire de Paris, DAEC,
              UMR 8631, CNRS et Universit\'e Paris 7,
	      F-92195 Meudon Cedex, France. \and
	      Laboratoire d'Astronomie Spatiale du CNRS, BP 8, 
	      Traverse du Siphon, F-13376 Marseille Cedex 12, France. \and
	      Visiting Astronomer, Kitt Peak national Observatory, National 
	      Optical Astronomical Observatories.}

   \date{Received / Accepted}

   \authorrunning{Vollmer et al.}

   \maketitle

\begin{abstract}
We present new H{\sc i} (20$''$ resolution) and CO observations of NGC 4548, 
an anemic galaxy in the Virgo cluster. 
The atomic gas distribution shows a ring
structure which is distorted at the northern edge. 
The overall rotation curve is derived with a tilted ring model. 
We compare our
rotation curve with previous ones and discuss the differences. 
The velocity field of the CO pointings fit very well with the one of 
the atomic gas where they overlap.
The CO emission permits to extend the rotation curve towards the galaxy 
centre. The molecular fraction is derived for the inner $120''\times 120''$
centered on the galaxy.
We compare the H{\sc i} and CO emission to H$\alpha$ line and
optical blue emission maps. The bar in the centre 
favors star formation at the outer end of the bar.
The H{\sc i} intensity distribution and velocity field
of the northern perturbation are deprojected with the help of a
first order kinematical model. They
are discussed in the framework of warps. It is concluded that the
scenario of ram pressure stripping responsible for the gas removal and
the northern perturbation is a very probable one. In this case
the ram pressure which may have caused the galaxy's H{\sc i} deficiency 
is now fading after the galaxy's close passage to the cluster centre or
increasing again due to a second approach of the galaxy to the 
cluster centre.
\end{abstract}

\keywords{
Galaxies: individual: NGC 4548 -- Galaxies: interactions -- Galaxies: ISM
-- Galaxies: kinematics and dynamics
}

\section{Introduction}

A galaxy cluster is an ideal laboratory for studying the influence 
of the galaxy's environment on its appearance and/or evolution. 
There are three types of interaction which can provide the mechanisms
for changing the galaxy's properties, i.e. morphology, 
luminosity and gas content as they enter the cluster. 
\begin{itemize}
\item
galaxy-galaxy gravitational interactions 
\item 
galaxy-cluster gravitational interactions due to the cluster potential
\item
galaxy-cluster interaction due to an individual galaxy's motion in the 
hot plasma of the cluster (ram pressure stripping).
\end{itemize}
In the case of galaxy-galaxy interactions one can observe important 
distortions of the stellar and gas content 
(see e.g. Barnes \& Hernquist 1996,
Combes 1997, Moore et al. 1996, Olson \& Kwan 1990) as the
tidal forces act on both components of the galaxy.
The influence of the cluster potential can cause distortions
of the stellar and gas content and its velocity field only
if the galaxy passes near enough the cluster centre (Valluri 1993).
However, ram pressure stripping (Gunn \& Gott 1972)
is only acting on the gas content. This provides
a tool to discriminate between these kinds of interactions
(Combes et al. 1988).\\
Spiral galaxies located near the cluster centre show very
different gas characteristics in a cluster than in the field. 
They are very deficient in H{\sc i} 
(Chamaraux et al. 1980, Bothun et al. 1982, 
Giovanelli \& Haynes 1985, Gavazzi 1987, 1989) 
and their H{\sc i} disk sizes
are also considerably reduced (van Gorkom \& Kotanyi 1985, Warmels 1988, 
Cayatte et al. 1990, 1994). 
Concerning the stellar content, their intrinsic colour indices
are not significantly different from field galaxies of the same
morphological type (Gavazzi et al. 1991, Gavazzi et al. 1998).
There is a special class of galaxies with a very low
arm inter-arm contrast defined as anemics by van den Bergh (1976).
But despite the H{\sc i} deficiency, cluster galaxies do not show a
reduced CO content (Kenney \& Young 1986, Casoli et al. 1991, 
Boselli et al. 1997a) neither
a reduced infrared luminosity (Bicay \& Giovanelli 1987).\\
A promising way to study the galaxy's interaction with its environment
is to look at details in the emission of the interstellar gas and its 
velocity distribution. 
As the atomic gas is located at large galactic radii and is relatively diffuse,
it is the most sensible tracer for perturbations which 
are induced by forces exterior to the galaxy.\\
The most common distortion of the outer H{\sc i} content of spiral galaxies
in general are warps. They can be described by a collection of spinning,
concentric rings whose angular moment vector is more and more inclined
with increasing radius (Rogstad et al. 1974). Thus, the un-tilted warp is 
a completely symmetric feature. Warps are a common feature in spiral 
galaxies (see e.g. Sancisi 1976, Bosma 1978, and Briggs 1990)
although their origin is still not clear (see e.g. Binney 1992, Jiang \& 
Binney 1999).
\begin{itemize}
\item
It can be interpreted as a discrete bending mode in flattened halos 
(Lynden-Bell 1965, Sparke 1984, Sparke \&  Casertano 1988).
However, if the proper dynamics of the halo are taken into account, the
warp fades away rapidly (Nelson \& Tremaine 1995, Binney et al. 1998).   
\item
According to cosmology galactic halos should be disturbed (Ryden \& Gunn 1987,
Ryden 1988, Quinn \& Binney 1991). They are
not in dynamical equilibrium and could so generate a warped disc.
\item 
Infalling galaxies could tidally induce or form a warp or a polar ring
(Jiang \& Binney 1999).
\item
Non-gravitational forces as ram pressure or magnetic pressure could
be responsible for a warp.
\end{itemize}
Binney (1992) points out, that ram pressure stripping should cause
an axisymmetric response which has the form of a rim.\\ 
In this paper, we present and discuss a new detailed H{\sc i} map of the 
anemic galaxy NGC 4548 which is located at 2.4$^{\rm o}$ from
the centre of the Virgo cluster looking at the kinematics in detail. 
The H{\sc i} map is compared to an H$\alpha$
line map and an optical B image. CO pointings with velocity information
are added to the cube in order to give more informations about
the kinematics at small galactic radii.  
A rotation curve is fitted and compared to the ones obtained by 
Guhathakurta et al. (1988) and Rubin et al. (1999). 
The local perturbation observed is discussed
in the framework of warps.\\

\section{Observations and data reduction}
\subsection{The H{\sc i} data}

The observations were made with the NRAO's Very Large Array (VLA)
\footnote{The National Radio Astronomy Observatory is operated by
Associated Universities Inc under a cooperative agreement with the
National Science Foundation.}, for 
description see Napier et al. (1983). The field was centered on NGC 4548
(Table 1).
We observed in December 20 1994 for 235 minutes with the C 
configuration and in
March 29 1995 for 90 minutes with the D configuration. A Hanning smoothing
was applied on-line to the initial frequency channels, yielding 
63 channels covering a total velocity width of 630 km\,s$^{-1}$. 
The velocity channels are centered on v$\sim$500 km\,s$^{-1}$.
The data were calibrated using the standard VLA reduction programs (AIPS).
A CLEANed image of all strong continuum sources were made and afterwards
directly subtracted from the UV data cube. At the end a linear interpolation
of the UV data points with respect to the frequency channels using the first
and last 10 channels was made in order to subtract the continuum.
The resulting image was CLEANed with a 20$''\times$20$''$ FWHM beam.
We ended up with a r.m.s. noise of $\sigma$=0.4 mJy/beam in one 
10 km\,s$^{-1}$ channel, or $\sigma$=4.7$\times 10^{19}$ cm$^{-2}$
expressed in column density.
\begin{table}
      \caption{The parameters of NGC 4548.
	 Col. (1) and (2): 1950 celestial 
 	coordinates. Col. (3): morphological type 
	(de Vaucouleurs et al. 1973). Col. (4): 
	heliocentric velocity, in km\,s$^{-1}$.
	Col. (5): H{\sc i} flux, in Jy\,km\,s$^{-1}$.
	Col. (6): H{\sc i} deficiency (Cayatte et al. 1994).}
         \label{TuAn1}
      \[
         \begin{array}{lccccc}
            \hline
            \noalign{\smallskip}
            $$\alpha$$(1950) & $$\delta$$(1950) & type & v$$_{\rm hel}$$ & 
{\rm HI\ flux} & {\rm HI\ def} \\
            \noalign{\smallskip}
            \hline
            \noalign{\smallskip}
            12h32m55.10s & 14$$^{\rm o}$$46$$'$$20.0$$''$$ & SBb(sr) & 486 & 9.4 & 0.77\\
            \noalign{\smallskip}
            \hline
         \end{array}
      \]
\end{table}
\subsection{The CO data}
\begin{figure}
	\resizebox{\hsize}{!}{\includegraphics{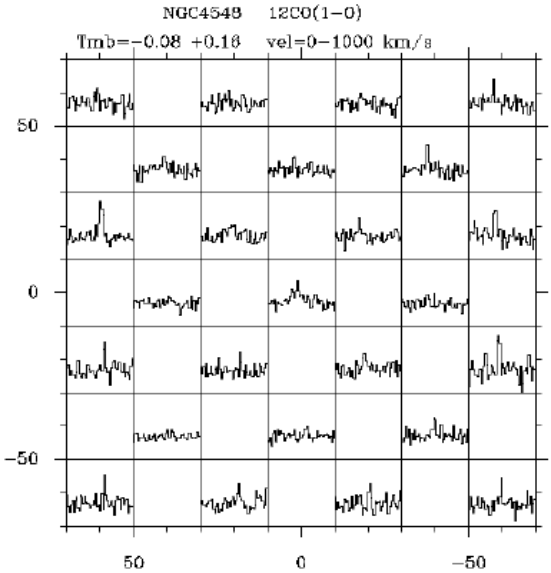}}
	\caption{The $^{12}$CO(1--0) spectra plotted on a $\alpha$-
	$\delta$ map centered on the nucleus of NGC 4548. The map's
	axes are offsets in arcseconds from this center. The velocity
	range of each spectrum is 0--1000 km\,s$^{-1}$ and the brightness
	temperature range is -0.08--+0.16 K.}
	\label{fig:COdata}
\end{figure}

NGC 4548 was observed in 1994 with the 
IRAM 30 m telescope at Pico Veleta (Granada, Spain). The beam size of the
telescope is 22$''$ at 115 GHz [$^{12}$CO(J=1--0)], 
which corresponds to 1.8 kpc at a distance of 17 
Mpc. Weather conditions were good, with typical zenith 
opacities of 0.25-0.45. 
The pointing accuracy was checked hourly by broad band continuum observations 
of the nearby source 3C273; the average error was 3$''$ rms. 
We used a SIS receiver in single sideband mode with T$_{rec}$=140-270 K  
and T$_{sys}$=500-800 K (in T$^*_A$ scale) at the elevation of the source. 
Two filter banks of 512 contiguous 1MHz channels provided a velocity 
resolution of 2.6 km s$^{-1}$ and a total velocity coverage of 
1330 km s$^{-1}$. We used a 
wobbler switching procedure, with a wobbler throw of 4$'$ in azimuth. Each 8 
minute scan began by a chopper wheel calibration on a load at ambient 
temperature and a cold load. The total integration time on each position was 
between 16 and 54 minutes on+off (i.e. half time on source) depending 
on the intensity of the signal, yielding a rms noise level of 15-30 mK 
(in the T$_{mb}$ scale) after boxcar velocity smoothing 
to 20.8 km\,s$^{-1}$.\\
NGC 4548 was observed at the nominal central coordinates and at 
different position offsets of 40$''$ (Fig.\ref{fig:COdata}). 
The galaxy was observed in 25 different positions 
and detected in 23 of them in the CO(1-0) line.\\
The data were reduced 
with the CLASS package (Forveille et al. 1990). The baselines 
were generally flat owing to the use of the wobbler, allowing us to subtract 
only linear baselines. The antenna temperature (T$^*_A$) was corrected for 
telescope and atmospheric losses by the calibration procedure. We adopt the 
main beam scale T$_{mb}$=T$^*_A$/$\eta$$_{mb}$ 
for the antenna temperature, with 
$\eta$$_{mb}$=0.56 for the CO(1-0) line. The integrated emission 
is given by: I(CO)=$\int$T$_{mb}$dv K km s$^{-1}$.

\subsection{The H$\alpha$ data}
The H$\alpha$ image (Fig.3) was obtained by J.R. Roy and P. Martin with 
the 1.6m of Observatoire du Mont
M\'egantic employing a f/8 $\rightarrow$ f/3.5 focal reducer, in May 1988.
The exposure time was 5 times 2000 sec with a 6577/10 filter. The FWHM
of the filter ($\sim$10\AA = 457 km\,s$^{-1}$) covers the whole velocity range
of the H{\sc i} data. NGC 4548 was observed with a RCA chip 360 
$\times$ 512 pixels (scale of 1.1 arcsec/pix).\\
The images where reduced using the software package IRAF following
the procedures described in Belley \& Roy (1992) and
in Martin \& Roy (1992). Two sources of uncertainties
relevant to the present analysis must be pointed out. First
the accuracy of monochromatic flux measurement in the inner regions
depends on how well the relatively bright stellar continuum 
in the central regions is subtracted;
the final result is somewhat uncertain because the continuum
filter used is about 400 \AA\ to the red side
of H$\alpha$. A first order  subtraction is usually done
using relative scaling of the flux of several stars in the field.
The resulting monochromatic image is that where the base level H$\alpha$
flux in interarm
regions (apart the presence of interarm H{\sc ii} regions)
shows net zero flux in the monochromatic image; 
this is achieved after careful
steps of trial and error.  Previous to this, 
subtraction of the sky background has been done.
This brings up the second source of uncertainties, which is
the accuracy of the flat fielding procedure.  This is quite critical
because it affects the measurements of mean H$\alpha$ surface
brightness in the faint outer regions. Uncertainties arise 
because of the restricted field and of the difficulties  of 
eliminating large-scale variations of the the 
background illumination to better
than 2-3 \% in monochromatic imagery done with
our focal reducer.  To circumvent this, we eliminated
the regions of weak surface brightness by  applying conservatively a 
high cut-off to the data. We kept only {\it individual} pixels which had
number of counts (in ADU) greater than the rms value above the background
in the annulus were the total number of counts was not longer seen to increase
with increasing galactocentric distance.\\
Finally, the images were calibrated using the aperture 
spectrophotometry done by McCall (1982).

\subsection{The optical data}
A CCD B image of NGC 4548 (Fig.2) was obtained at the NOAO 0.9 m 
telescope at Kitt 
Peak\footnote{Kitt Peak national Observatory, National Optical Astronomical 
Observatories, which are operated by the Association of Universities for
Research in
Astronomy, Inc. (AURA), under cooperative agreement with the National Science
Foundation.}, Arizona, in June 1995. We used a 2048x2048 T2KA detector 
(2 e$^-$/ADUs) 
in the f/13.5 configuration, which gives a field of view of $13.1'\times 13.1'$ 
with a pixel size of 0.4$''$. 
The galaxy was observed through the Kitt Peak B Harris 
filter in a 15 min exposure during non photometric conditions. The image 
was reduced using standard procedures, including bias correction, dome flat 
fielding and cosmic ray removal. The seeing was $\sim$2$''$. 

\section{Results}

\begin{figure}
	\resizebox{\hsize}{!}{\includegraphics{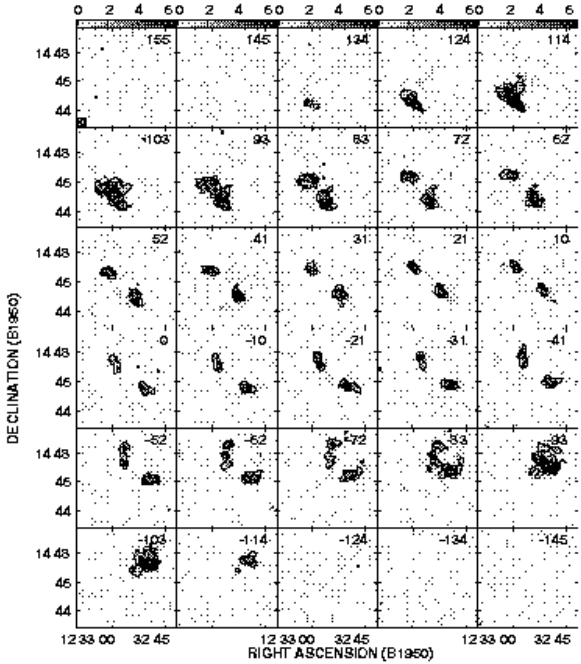}}
	\caption{The H{\sc i} channel maps. The heliocentric velocity
	relative to the reference velocity
	is indicated in km\,s$^{-1}$ in the upper right of each channel.
	The contour lines are 1.3, 2.6, 3.9, 5.2, and 6.5 mJy/beam.}
	\label{fig:PBCOR}
\end{figure}

\begin{figure}
	\resizebox{\hsize}{!}{\includegraphics{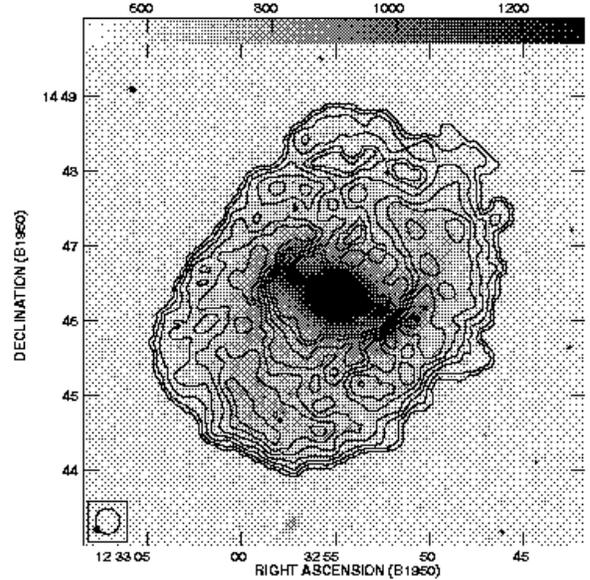}}
	\caption{The optical B image together with the H{\sc i} 
	contour map of NGC 4548. The contour levels
	correspond to 0.82, 3.24, 6.48, 9.73, 12.97, 16.21, 
	19.45$\times 10^{20}$ cm$^{-2}$. The beam is shown
	in the lower left corner. In order to distinguish
	maxima and minima we refer to Fig.5.}
	\label{fig:BLUECO}
\end{figure}

\begin{figure}
	\resizebox{\hsize}{!}{\includegraphics{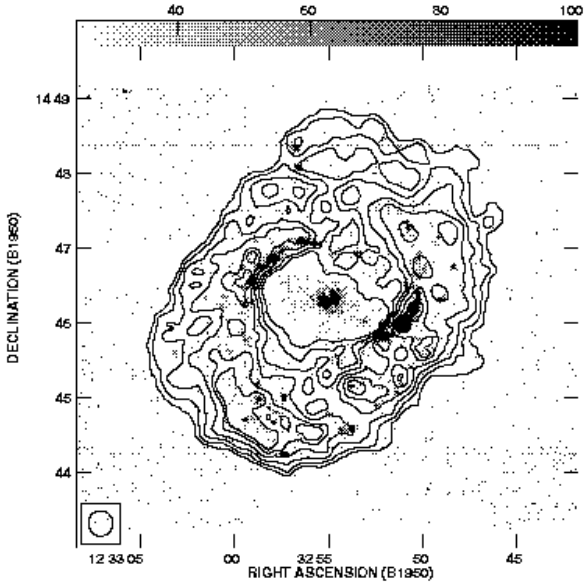}}
	\caption{Contour plot of the H{\sc i} emission map
	together with the H$\alpha$ image. The contour levels
	correspond to column densities of 3.44, 6.89, 10.33, 13.78, 17.22, 
	19.29$\times$10$^{20}$ cm$^{-2}$. The H{\sc i} beam is shown in 
	the lower left corner.
	\label{fig:HACO}	
	}	
\end{figure}

\begin{figure}
	\resizebox{\hsize}{!}{\includegraphics{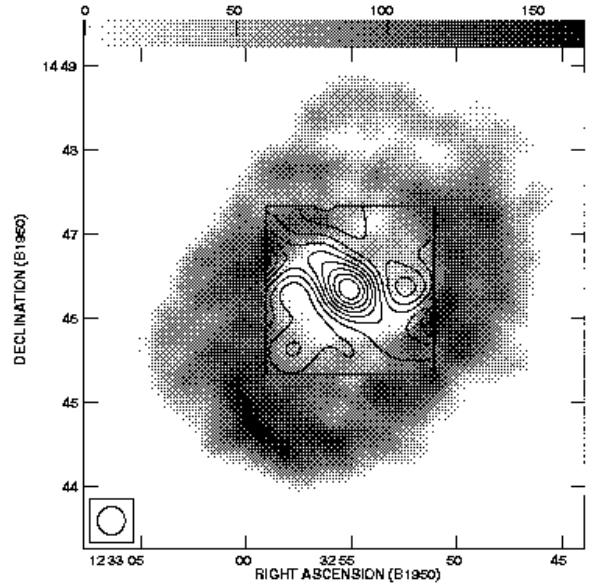}}
	\caption{The H{\sc i} emission map from Fig.\ref{fig:BLUECO} together 
	with CO emission contour map. The maximum level 
	is 10.87 K\,km\,s$^{-1}$.
	The contour levels are in steps of 1.09 K\,km\,s$^{-1}$.
	The H{\sc i} beam is shown in the lower left corner.
	\label{fig:TOTCO}
	}	
\end{figure}

\begin{figure}
	\resizebox{\hsize}{!}{\includegraphics{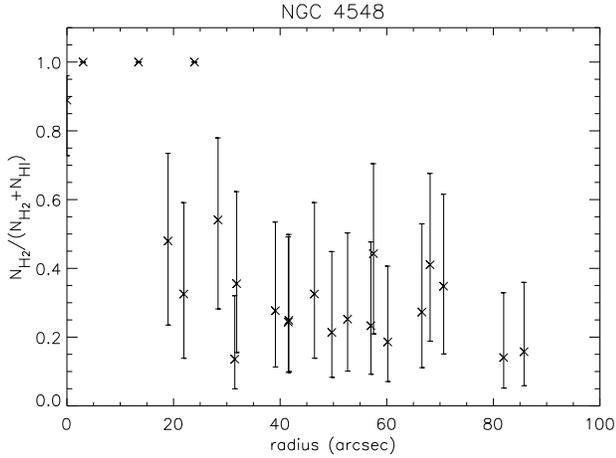}}
	\caption{The molecular fraction at each CO pointing.
	Its values as a function of the deprojected distance to the 
	galaxy centre are shown. The error bars correspond
	to conversion factors of $X/3$ and $X$$\times$$3$.
	} \label{fig:HtoCO}
\end{figure}

The complete H{\sc i} data cube is shown in Fig.\ref{fig:PBCOR}.
The flux added over all channels with a cutoff of 2$\sigma$ is shown in 
Fig.\ref{fig:BLUECO}. 
The total flux calculated using a cutoff of 2$\sigma$ is S$_{{\rm HI}}$=
9.4 Jy\,km\,s$^{-1}$. This corresponds to a total H{\sc i} mass of
M$_{{\rm HI}}=6.4\times 10^{8}$ M$_{\odot}$
with a distance D = 17 Mpc for the Virgo cluster \footnote{We assume 
H$_{0}$=75 km\,s$^{-1}$Mpc$^{-1}$.}. This is in good agreement with 
the value given by Cayatte et al. (1990) using a larger beam size of 
$55''\times 55''$.\\
The H{\sc i} emission is distributed within an almost complete ring.
The maximum of emission is located in the south-east. The emission
profile along the minor axis is quite symmetrical, whereas the one
along the major axis is strongly asymmetrical. In fact the south-eastern
emission maximum has no counterpart in the north-west. 
The near Infrared image of Boselli et al. (1997b Fig.3b) shows  
clearly the bar and the two spiral arms in the north-west and south-east.
As for the H{\sc i} emission, there is more NIR emission
coming from the southern part of the galaxy. In addition, 
the outer limits of the NIR emission coincide well with the one
of the H{\sc i} emission.
Thus, the intensity of the H{\sc i} emission follows the gravitational
potential traced by the NIR image.
In the galaxy's centre the H{\sc i} emission drops by an order of 
magnitude leading to
an in east-west direction elongated hole. It is also worth
noticing that the inner edge of the emission ring 
extends more inwards in the north than in the south.\\
In order to compare the galaxy's gas content to its stellar population,
we show the H{\sc i} emission together with an optical B image 
in Fig.\ref{fig:BLUECO}.
One clearly recognizes the bar which ends at the inner edge of the
H{\sc i} ring. It is also visible that both the low contrast spiral arms
are traced by the H{\sc i} emission. The local H{\sc i} emission
maximum in the extended southern arm is associated with the young stellar
population traced by the spiral arm. The outer edge of the
atomic gas ring follows exactly the shape of the stellar disc. 
A dust lane is seen in absorption near the centre in the south-west. 
If one accepts the idea that it is not located in the inner disc, 
this indicates that
the eastern side is the near side of the galaxy.\\
As expected for an anemic spiral galaxy the H$\alpha$ line map shows
very few H{\sc ii} regions. In Fig.\ref{fig:HACO} we show this map together with
the H{\sc i} emission map.
The most luminous H{\sc ii} regions are located
along the beginning of the spiral arms at the end of the bar. 
There the interaction between the bar and the outer gas favours
star formation.\\
The CO emission was observed in a $120''\times120''$ region
centered on the galaxy. These data together with the H{\sc i} map
are shown in Fig.\ref{fig:TOTCO}.
As expected, the bar appears clearly in the CO emission. 
We can also observe the points where the local CO emission
maxima join the H{\sc i} emission maxima at the ends of the bar.\\
Honma et al. 1995 demonstrated that the gas phase transition 
between H{\sc i} and H$_{2}$ occurs within a small radial distance.
The fraction of H$_{2}$ to the total gas column density (molecular fraction)
increases very rapidly inwards within this boundary region. 
In order to study this effect for NGC 4548, the deprojected 
distance for each CO pointing to the galaxy centre was calculated 
with the help of the position angle and the inclination (see next section).
As the H{\sc i} and CO data have similar beam sizes it is possible
to compare the column densities at a given position.
The fraction
of column densities $x=N_{H_{2}}/(N_{H_{2}}+N_{HI})$ for each CO pointing
is plotted as a function of the deprojected distance (Fig.\ref{fig:HtoCO}). 
We assumed CO conversion factor of $X = 1.0\pm 0.1\, 10^{20}$ cm$^{-2}$ 
(K km s$^{-1})^{-1}$ as derived by the EGRET gamma-ray observations
(Digel et al. 1996) for the solar neighbourhood. However, in the Perseus arm 
at 3-4 kpc from the Sun $X=2.5\pm 0.9\, 10^{20}$ cm$^{-2}$ 
(K km s$^{-1})^{-1}$ (Digel et al. 1996). For extragalactic sources there
are only estimations ranging from $X\simeq 0.6\, 10^{20}$ cm$^{-2}$ 
(K km s$^{-1})^{-1}$ for M51 (Gu\'elin et al. 1995) to 
$X\simeq 10\, 10^{20}$ cm$^{-2}$ (K km s$^{-1})^{-1}$ for the 
SMC (Lequeux et al. 1994). In general it seems that 
$X$ does not differ by  a large factor in spiral galaxies 
with a luminosity similar to the Galaxy (Boselli et al. 1997a) .
Therefore, we have adopted a factor 3 for the uncertainties in the 
determination of $X$. The error bars in Fig.\ref{fig:HtoCO} represent 
these uncertainties.
As there is no H{\sc i} detection in the centre $x=1$ there.
The molecular fraction tends to decrease with radius up to $\sim 40''$ 
which corresponds approximately to the 
radial extent of the bar. There the predominantly atomic gas appears to 
be transformed into molecules due to the compression caused by the bar.
Further out the molecular fraction does not show a further decline.
This means that we observe a sharp transition between molecular and 
atomic gas at about 30$''$ and a constant molecular gas fraction 
$x=0.27\pm 0.09$ (assuming $X=10^{20}$ cm$^{-2}$ (K km s$^{-1}$)$^{-1}$)
further out. 

\section{Kinematics}
\subsection{The rotation curve}

\begin{figure}
	\resizebox{\hsize}{!}{\includegraphics{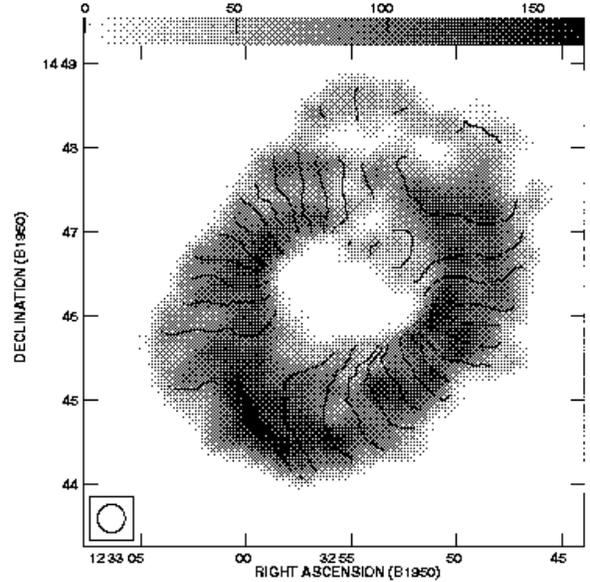}}
	\caption{The H{\sc i} emission map together with the contours
	 of the intensity weighted mean velocity field.
	The peak contour level is at 602 km\,s$^{-1}$ in the south-east
	of the galaxy centre. The contour levels are in steps of 
	15 km\,s$^{-1}$ northwards. The H{\sc i} beam is shown in 
	the lower left corner.	
	} \label{fig:TOTNEW}
\end{figure}

\begin{figure}
	\resizebox{\hsize}{!}{\includegraphics{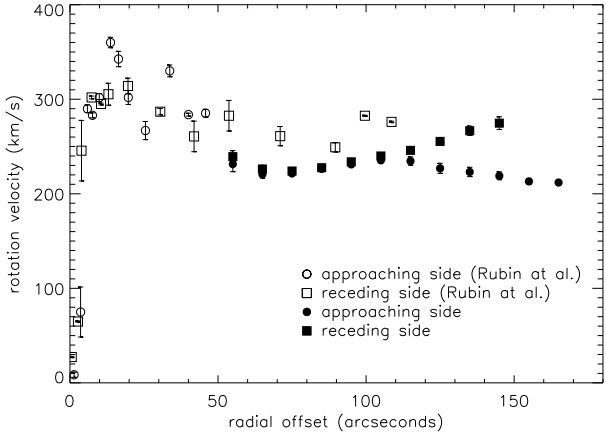}}
	\caption{The rotation curve of NGC 4548. The open boxes and
	circles are the Rubin et al. (1999) data. The filled boxes and 
	circles represent our data. The bars indicate the errors on the 
	measurements. If they are not visible, they do not extend further
	than the symbol size.
	} \label{fig:rotat}
\end{figure}

The H{\sc i} emission map together with the contours of the intensity 
weighted mean velocity field is plotted in Fig.\ref{fig:TOTNEW}. 
The overall velocity field is in good agreement with an almost
unperturbed rotation around the galaxy centre.
Nevertheless, we can see a distortion in the south-east caused
by the spiral arm which is very prominent in H{\sc i} there.
A minor distortion can also be seen in the north-west, where the
other spiral arm is located. There are first hints that the
northern part of the galaxy has a somewhat peculiar velocity field which
does not fit the rest.\\
The position angle and the inclination were determined using a tilted ring 
model (see Begeman 1987). 
We have used the optical central position (Table 1) and a
systemic velocity of 475 km\,s$^{-1}$ (Rubin et al. 1999) as initial values.
We averaged first the approaching and receding side of the galaxy
up to a radius of 115$''$ excluding points within a sector of 
$\pm 30^{\rm o}$ around the minor axis.
The derived dynamical centre is the optical one $\pm 3''$.
The results for the other parameters are given in Table 2.\\
 \begin{table}
      \caption{Col. (1) Author. Col. (2): Position angle of the major axis. 
	Col. (3): 
	inclination angle. Col. (4): systemic velocity in km\,s$^{-1}$.
	Col. (5): maximum rotation velocity in km\,s$^{-1}$.}
         \label{TuAn2}
      \[
         \begin{array}{lcccc}
            \hline
            \noalign{\smallskip}
            {\rm author} & {\rm position\ angle} & {\rm inclination} & v_{sys} & v_{max} \\
            \noalign{\smallskip}
            \hline
            \noalign{\smallskip}
            {\rm Vollmer} & 136$$\pm 0.5^{\rm o}$$ & 25$$\pm 6^{\rm o}$$ & 477$$\pm 1$$ & 254$$\pm 47$$ \\
	    {\rm Guhathakurta} & 135$$^{\rm o}$$ & 42$$^{\rm o}$$ & 504 & 200 \\
	    {\rm Rubin} & 136$$^{\rm o}$$ & 38$$^{\rm o}$$ & 475 & 200 \\
            \noalign{\smallskip}
            \hline
         \end{array}
      \]
\end{table}
The systemic velocity is in excellent agreement with the value of Rubin 
et al. (1999) and agrees well with the value given by Cayatte et al. (1990),
whereas the value given by Guhathakurta et al. (1988) differs by $\sim 5\%$.
The position angle is in excellent agreement with Rubin et al. (1999)
and Guhathakurta et al. (1988). The main difference between the
two authors and us is the value of the inclination angle. Our value is about 
10-15$^{\rm o}$ lower than the ones previously derived.
This is due to the complex structure of the rotation curve which we 
will discuss now. 
In a second step we fixed the dynamical centre and the systemic velocity
of the tilted ring model and fitted the approaching and receding side
separately. The resulting inclination angle increases for both sides
with increasing radius. The rotation velocity stays constant in the
south-eastern part of the galaxy. On the north-western side it decreases
up to 140$''$ and then increases with increasing radius.
In order to compare our rotation curve with the one of Rubin et al. (1999)
we fixed the inclination angle at 27$^{\rm o}$ for both data. 
This choice gives coherent rotation curves.  
Both rotation curves are shown in Fig.\ref{fig:rotat}. 
It is best fitted by a constant rotation velocity of $\sim$250 km\,s$^{-1}$.
For radii greater than 120$''$ the rotation curves of the two sides
diverge. 
The nearly constant slope in the rotation curve of the receding side
in our data suggests that this is due to a rising inclination angle.
In order to investigate on the change of the inclination we fixed the
rotation velocity at 250 km\,s$^{-1}$. In this case the inclination angle has
has the same behaviour as the rotation curves in Fig.\ref{fig:rotat}.
It has the same value up to a radius of $\sim 110''$ and diverges for 
greater radii.
\begin{figure}
	\resizebox{\hsize}{!}{\includegraphics{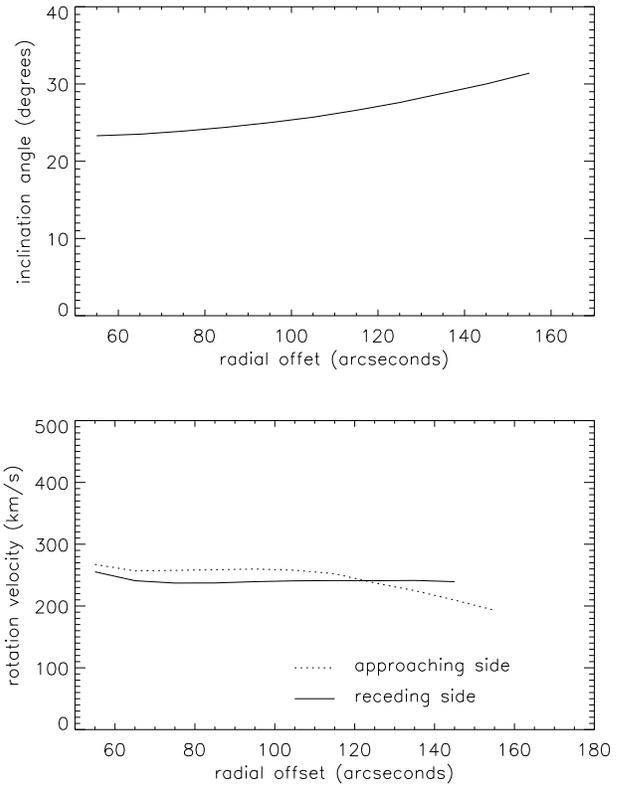}}
	\caption{Top: Inclination angle of both sides 
	with respect to the radial offset
	for case (iii). Bottom: The resulting H{\sc i} rotation curves
	for the approaching and receding side.
	} \label{fig:rotat2}
\end{figure}
The difference in the shape of the rotation curve with the
one derived by Guhathakurta et al. (1988) is due to the fact that 
their data had twice the beam size and
five times less sensitivity. The worse angular resolution caused
major beam smearing effects and the low sensitivity permitted only
to fit the rotation curve to a smaller part of the H{\sc i} emission 
revealed by our observations. 
The consequence is that they derived a more steeply rising rotation curve 
with increasing radius for both sides.\\
The divergence of the rotation curve for radii greater than 120$''$
could be due to three extreme cases. (i) The rotation velocity is
constant and the inclination angle decreases at the western side and
increases for the eastern side. (ii) The inclination angle stays 
approximately constant for both sides.
In this case the neutral gas is accelerated in the east and decelerated
in the west. (iii) The inclination angle increases
for both sides and the gas at the western side is decelerated.
A tilted ring model where the rotation curves are
fitted separately for each side using the same increasing inclination angle
(Fig.\ref{fig:rotat2})
gives residuals which are approximately 3 times less than those of the
previous models. Therefore, this hypothesis seems to be the most
probable.

\subsection{A three dimensional method to visualize the H{\sc i} data cube}

\begin{figure}
	\resizebox{\hsize}{!}{\includegraphics{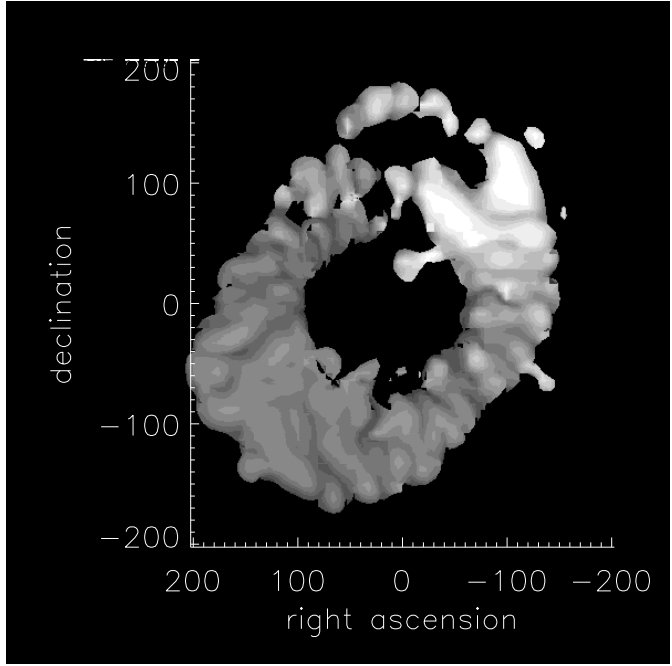}}
	\caption{Three dimensional visualization of the data cube with
	an intensity level of 2 mJy/beam. The brighter the structures
	are the nearer they are to the observer.
	} \label{fig:paris1}
\end{figure}

\begin{figure}
	\resizebox{\hsize}{!}{\includegraphics{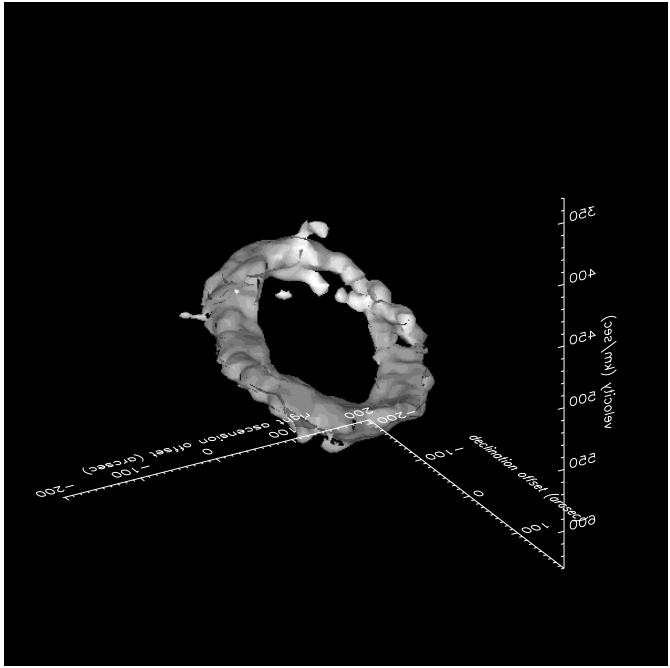}}
	\caption{The same cube as in Fig.\ref{fig:paris1} but rotated by
	$\sim$180$^{\rm o}$ around the heliocentric velocity axis.
	} \label{fig:paris3}
\end{figure}

In order to have a more detailed and complete view of the kinematics, it
is necessary to look at the data cube as a whole in three dimensions
to separate connected features more clearly
\footnote{The visualizations are done with IDL (version 5.1).}.
All velocity channels seen in Fig.\ref{fig:PBCOR} 
are piled up to give the cube. The cube's axes are right ascension, 
declination and heliocentric velocity. All points in the cube having 
intensities exceeding a chosen level become opaque, the rest
being transparent. The surface created in this way is illuminated
by light which is coming out of the observer's direction.
Thus, brighter features are closer to the observer.
This representation allows to analyse the whole cube from any
possible point of view. 
The data cube visualized in this way with
an intensity level of 2 mJy/beam (5$\sigma$) is plotted in Fig.\ref{fig:paris1}.
The z-axis is perpendicular to the image plane.
It corresponds to Fig.\ref{fig:BLUECO} with
the important difference that this is a three dimensional
representation where the third axis is perpendicular to the
image plane. In order to examine the detached arm, we turn the cube 
$\sim$180$^{\rm o}$
around the heliocentric velocity axis. This can be seen in 
Fig.\ref{fig:paris3}.
In this representation the constant rotation velocity translates
into a ring in the three dimensional restricted phase space 
($\alpha$, $\delta$, heliocentric velocity). The most interesting 
things happen in the north, where the detached arm described above can be
clearly distinguished.\\
It has the shape of a filament `warped' around the constant 
velocity ring in the restricted phase space. 
The emission blob at the top corresponds to the emission blob
in Fig.\ref{fig:paris1} which reaches into the central hole from 
the north-west. It is the emission nearest to the galaxy centre. 
The complexity of the  rotation curve described above can be
observed directly as a discontinuity of the three dimensional
velocity field in the north.

\subsection{Adding the CO pointings to the cube}

\begin{figure}
	\resizebox{\hsize}{!}{\includegraphics{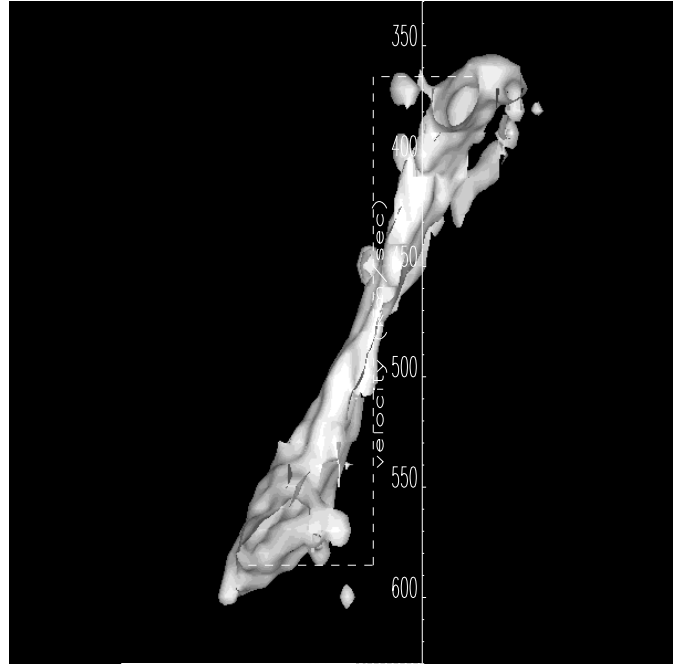}}
	\caption{The constant velocity rotation ring seen edge-on.
	The $\alpha - \delta$ plane is perpendicular to the image
	plane, the velocity axis is seen vertically.
	The case of ideal constant velocity rotation is shown
	by the broken line.
	} \label{fig:paris12}
\end{figure}

\begin{figure}
	\resizebox{\hsize}{!}{\includegraphics{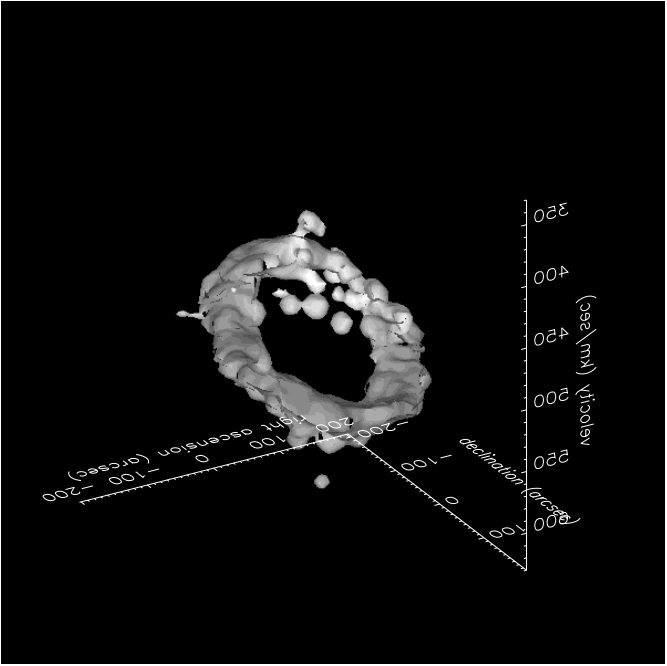}}
	\caption{The same view as in Fig.\ref{fig:paris3} but with the CO
	points included. The spheres within the constant velocity
	rotation ring are the pointings which lie on the bar.
	The CO points located within the H{\sc i} ring are nicely
	embedded in the H{\sc i} emission.
	} \label{fig:paris13}
\end{figure}

The 23 points can be easily included into the data cube.
Because of the sparseness of the CO detections it is
not possible to apply an interpolation in order to have a 
continuous velocity field.
As the H{\sc i} cube has a quasi continuous nature whereas we have
only 23 discrete points in CO, we have to adjust the CO intensity
and smooth it in the following way. In order to have the best view
of the ensemble we decided to normalize the maximum CO intensity
to that of the H{\sc i} intensity. Then each CO data point
is smoothed with a Gaussian filter with a FWHM of 20$''$.
The CO pointings are now spheres in the restricted phase space 
whose radii are comparable to
the beam size which are easily visible in contrast
to the surrounding H{\sc i} features. It should be stressed
that the aim of this procedure is to compare H{\sc i} and CO velocities,
disregarding any flux conservation.\\ 
Now we have to investigate if the CO points which lie inside
the contours have the same velocities as their H{\sc i} counterpart.
Furthermore, making a single cube of both data will permit us
to see if there is a smooth junction in the restricted phase space
between the spatially inner CO point and the outer H{\sc i} emission
structure. Fig.\ref{fig:paris12} shows the ring of constant velocity rotation
edge-on. 
The detached arm is located in the north. The $\alpha$-$\delta$
plane is perpendicular to the image plane, the heliocentric velocity axis
is seen vertically. With the help of this image two conclusions
can be drawn. (i) The CO points  indicate that the rotation
with approximately constant velocity extends further into the centre as the 
CO points fit into the ideal constant velocity field which is
sketched by the dashed line. Especially the H{\sc i} blob nearest to the 
galaxy centre described above has a CO pointing nearby which fits
perfectly into the constant velocity rotation scenario. 
(ii) The CO points match nicely
into the H{\sc i} data cube. No major discrepancies are detected
except a point which is located in the bar. But as this gas
is expected to be highly perturbed, this is not surprising.
In this figure one can easily recognize the greater velocities
as expected for a constant velocity rotation in the east and the
lower velocities compared to a constant velocity rotation in the west.
Finally, in order to show the nice smooth junction between both
data, Fig.\ref{fig:paris13} shows the same view as in Fig.\ref{fig:paris3} 
but including the CO points. Here one can
see the CO spheres embedded nicely in the H{\sc i} structures.  

\subsection{First order modeling}

\begin{figure}
	\resizebox{\hsize}{!}{\includegraphics{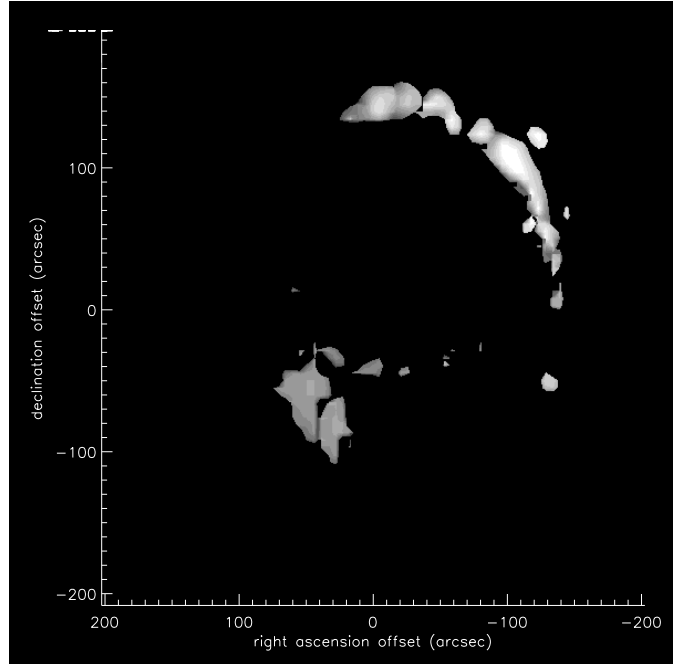}}
	\caption{Residual of the subtraction of the model cube
	from the H{\sc i} data cube. 
	} \label{fig:paris5}
\end{figure}

\begin{figure}
	\resizebox{\hsize}{!}{\includegraphics{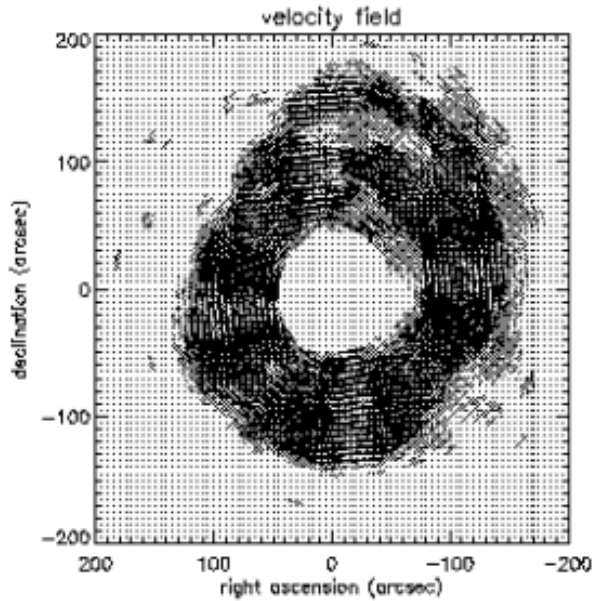}}
	\caption{The reconstructed sites of H{\sc i} emission.
	Dark: only velocity vectors of model points corresponding 
	to emission in the data cube above the given level are
	plotted. Grey: the intensity level in the H{\sc i} cube
	is lowered by a factor 2.
	} \label{fig:velovecti}
\end{figure}

As the velocity field from 50$''$ up to 100$''$ (projected radius) is known,
it is possible
to make a simple model and compare it directly with the cube.
We have distributed 10000 points in two dimensions between 50$''$
and an outer deprojected radius r$_{\rm out}$. A constant azimuthal velocity
of 250 km\,s$^{-1}$ was attributed to them. At the end a randomly
chosen velocity vector with a length of up to 10 km\,s$^{-1}$ 
(van der Kruit \& Shostak 1982) and
a randomly chosen direction in three dimensions was added to the velocity
of each point.
With this model velocity field a cube similar to  the one discussed above
can be constructed. As the given field is one single realization
of the model and we are interested in the general case, we smooth
the model cube with a second order Gaussian filter.
This corresponds - through hydrostatic equilibrium - to a certain
thickness of the H{\sc i} disk.
We then subtracted the model cube from the data cube. This procedure
gives us the emission structure which can not be explained by
a constant velocity rotation field ranging between 50$''$ (deprojected radius)
and r$_{\rm out}$. The outer radius is fixed by the following condition.
We begin with a small number, say 100$''$ and increase it as long
as there is a residual in the south east of the image (Fig.\ref{fig:BLUECO}).
This procedure gives us a radius of r$_{\rm out}$=135$''$.
The final residual is shown in Fig.\ref{fig:paris5}. 
As a result, the complete northern detached arm appears very clearly.
So this part of the galaxy is not located within the given radius
in the disk plane.
In addition, parts of the spiral structure in the south-east whose
velocity field differs from the constant velocity rotation are visible.\\
If we extend the outer deprojected radius in the model to 
r$_{\rm out}$=170$''$, even the northern arm can be recovered by the model.
The modeling applied allows us to determine easily which model points
correspond to emission points in the cube above the given level.
The deprojected distribution of sites of H{\sc i} is therefore accessible.
It is shown in Fig.\ref{fig:velovecti} as dark lines. 
The grey lines correspond to
an intensity level in the H{\sc i} cube which is half of the one
previously used. They correspond to the noise level in the cube.
The reconstructed low intensity extended emission velocity field
shows a well defined straight ridge from the eastern to the
northern edge. In addition, in the north the galaxy's outer H{\sc i} emission
extends further to the north than expected assuming a constant outer 
galactic radius.   
In the reconstruction using the higher intensity level the 
northern detached arm clearly appears. The detached blob which can
be seen in Fig.\ref{fig:paris1} at $\alpha \sim -120'', \delta \sim 120''$ is
recovered by the model too. It is quite remarkable that the upper
part of the northern arm has already here a well defined vertical edge.\\
The next stage in modeling is to include a simple warp. Placing
all model points exceeding r=135$''$ in a plane which 
is inclined by 15$^{\rm o}$ with respect to the galaxy's plane appeared 
to be a good fit to the data. 
In this case the northern arm is located in projection
behind the galaxy's plane. The distance between the northern arm
and the galaxy centre is thus bigger in this case than in the one
assumed above. Applying the same procedure as before,
gives similar results. The fact that we have only
emission at one side of the galaxy together with the
smallness of the tilt angle does not permit any
distinction between the two cases. Therefore the modeling
of a increasing inclination angle does not give very different results
for the reconstructed H{\sc i} emission distribution.\\
It should be stressed here that the aim of the modelization is not 
to reproduce the exact velocity field of the H{\sc i} data but
to show general effects. The conclusion which can be drawn is that
there is an asymmetrical structure in the deprojected disk which
can be fitted by a much increased outer radius of the galaxy
or by material which is located in tilted plane with respect to the 
galaxy's plane. This confirms the results of the tilted ring model.
Furthermore, we can see a well defined north-eastern ridge in the
reconstructed velocity field for the low intensity extended H{\sc i} 
emission in both models.

\section{Discussion}

We will now discuss the possibilities to
explain the complex structure of the velocity field in the 
restricted phase space.
\begin{itemize}
\item
If the galaxy's outer deprojected radius is $\sim 170''$, 
the whole northern arm with the detached blob at $\alpha \sim -120'' \ \ 
\delta \sim 120''$ is located within the plane of the disk. 
This represents the simplest case and is shown in Fig.\ref{fig:velovecti}.
In this case the neutral gas in the eastern part of the galaxy is
accelerated and the gas in the western part is decelerated.
\item
The northern arm could be located beyond the disk plane in which the galaxy
is located. The rotation velocity is still constant.
In this case the inclination angle diverges for radii greater than 120$''$. 
\item 
The rotation velocity field is distorted, but the gas moves in the
plane of the galaxy whose inclination angle increases with increasing
radius. This means that the material in the eastern part
is strongly decelerated (Fig.\ref{fig:rotat2}).
\item
The velocity field is distorted and the gas is not located in the
galaxy plane. This is the most complicated case and can only
be verified if a dynamical model is applied.
\end{itemize}
The fitting of the tilted ring model already gives an idea which
possibility is the most probable one.
The three dimensional representation completes the analysis giving
additional hints which possibility is preferable.
The two methods are thus complementary.
As Fig.\ref{fig:paris3} shows a clear discontinuity of the
velocity field in the north, we think that an increasing
inclination angle for both sides is most likely. This confirms
the conclusion derived with the tilted ring model. 
In addition, Fig.\ref{fig:velovecti} shows a `perturbed' H{\sc i} 
emission distribution and therefore atomic gas mass 
distribution (the detached northern arm and the north-eastern ridge).
This suggests that the same mechanism which produces the ridge
might be responsible for the complexity of the rotation curve and
especially the perturbation of the velocity field 
in the north. Thus, the constant velocity assumption is not
the whole story for the northern detached arm.\\
The near-IR H band image of NGC 4548 (Boselli et al. 1997b) which traces
fairly well the gravitational potential does not show a 
significant asymmetry
in the north which could be responsible for the disturbed velocity field. 
Therefore, an external force is necessary 
to cause and/or to maintain the north-eastern ridge and
the northern perturbation. NGC 4548 is located at a projected distance of 
2.4$^{\rm o}$ of the cluster centre (M87). If we suppose that
the deprojected distance is not significantly higher, the galaxy
is moving in a high temperature (T$\sim 10^{7}$ K) and low density
($\rho \sim 10^{-4}$cm$^{-3}$, B\"ohringer et al. 1994) 
gas at a speed of v$\sim$1000 km\,s$^{-1}$.
Cayatte et al. (1994) have shown that assuming these conditions the
ram pressure of the hot intra-cluster medium can exceed the restoring
force of the galaxy's gravitational potential. So it is very probable
that this external ram pressure force can explain the observed 
perturbation in the atomic gas content.\\
Nevertheless, we can not entirely exclude that this is a case of a warp not 
induced by ram pressure. H{\sc i} warps are a common feature of 
spiral galaxies (see e.g. Bosma 1981, Briggs 1990, Bottema 1995).
In their sample of 1700 galaxies 
Richter \& Sancisi (1994) found that 50\% of the galaxies
show asymmetries, non-circularities, or lopsidedness in the H{\sc i} 
distribution. Reshetnikov \& Combes (1998) showed that in their sample
of 540 galaxies half of them have optical warps.
The majority of H{\sc i} warps can be described by a tilted ring model,
where the position angle and the inclination change with radius for
the whole galaxy.
Compared to these warps, the distortion in the velocity field of NGC 4548
shows one important difference: the divergence of the inclination
angle and/or the rotation velocity.\\   
Other possibilities for the explanation of the distorted velocity
field are a perturbation of the halo,
a recent accretion of a dwarf galaxy, or a perturbation
of the gravitational field by one of the two nearby dwarf galaxies or by the 
gravitational potential of the cluster.
In all these cases the perturbation must have happened very recently
($t \sim 10^{8}$ yr) to show the observed asymmetry.\\
Instead, if we accept that the atomic
gas content is removed when the galaxy passes near the center
(Cayatte et al. 1994) this galaxy may have already passed the cluster once
as it is highly H{\sc i} deficient. So we observe either 
\begin{itemize}
\item
the fading effects of the ram pressure stripping if the galaxy is moving
away from the cluster centre,
\item
the increasing effects of a new stripping process in the opposite case, 
\item
the effects of a halo perturbation,
\item
the influence of the accretion of a dwarf galaxy 
\item
a mixture of these effects.
\end{itemize} 
Both perturbations due to gravitational interaction do not explain 
the north-eastern ridge neither the divergence of the inclination
angle / rotation velocity in a natural way.
On the other hand, ram pressure acts more efficiently on clouds
of small column density. Therefore it would be only natural to
observe its consequences on gas with a small column density
(which corresponds to a region of low gravitational potential) , i.e.
the detached northern arm and the north-eastern ridge.
This confirm the conclusion drawn above, i.e. that the galaxy's 
inclination angle rises for both sides and that the gas in the 
north-east is decelerated.
In addition, NGC 4548 has a radial velocity of about -800 km\,s$^{-1}$
with respect to the cluster centre (M87). If the eastern side
is in front of the western one as the absorption feature suggests,
the galaxy's motion is directed to the north-east. The pressure
due to the intra-cluster medium decelerates the neutral gas
of low column density. This deceleration is observed for
the detached northern arm (Fig.\ref{fig:paris12}).

\section{Conclusion}

We have shown a high sensitivity H{\sc i} data cube of NGC 4548 together
with 23 pointings in CO(1-0), an H$\alpha$ line map, and an
optical B image. 
The H{\sc i} emission shows a ring-like structure which is symmetric 
along the minor axis and asymmetric along the major axis. Particularly,
there is a distorted low intensity arm in the north. The comparison
of this data with an optical B image showed that the H{\sc i} emission
follows the spiral arms. The detached northern arm however, 
has no detectable counterpart in the B image. 
The dust lanes appearing in the south-west places
this side behind the eastern part of the galaxy.\\ 
The H$\alpha$ line emission shows several bright H{\sc ii} regions
at the ends of the bar.\\
The CO(1-0) data covers $120''\times 120''$ centered on the galaxy. 
The bar appears clearly  and joins the H{\sc i} emission at the end of the bar.
The molecular fraction stays constant for galactic radii greater than 40$''$
and rises rapidly between 20$''$ and 40$''$.\\
We have determined the position angle and the inclination and have fitted
a rotation curve to the atomic gas data. The velocity field can be described
either by a constant velocity rotation of $\sim$250 km\,s$^{-1}$ and
a diverging inclination angle for the outer parts or by a
constant inclination angle and a diverging rotation velocity for the
outer parts or by something in between. The difference in the shape
of the rotation curve
with the one derived by Guhartakurta et al. (1988) can be explained by 
the better angular resolution and the much better sensitivity of our
data.\\
We choose a three dimensional visualization of the data to show
the detailed velocity field. In three dimensions the constant velocity
rotation corresponds to a ring in the restricted phase space ($\alpha$,
$\delta$, heliocentric velocity). The detached northern arm is clearly visible
and it appeared that it is distorted in velocity too. The inclusion
of the CO data into the atomic gas cube gives insight into the kinematics
of the central region and fits nicely with the data of Rubin et al. (1999).\\
A simple model of a disk with constant rotation velocity is
fitted to the data. In the reconstructed emission map we observe
an asymmetry in the north. We list the different possibilities for
the causes of the distorted velocity field and discuss them in detail.\\
We conclude that the ram pressure 
stripping scenario is entirely consistent with all observational
constraints and might be the dominating effect.
If this is the case there might be the possibility to have a pile-up
of the intracluster medium
in front of the galaxy in the direction of it's motion.
X--ray observations with XMM detecting this pile-up may thus 
confirm our conclusions. Furthermore, we will investigate further on this topic
using detailed dynamical models of this scenario.
The galaxy's velocity with respect to the cluster centre (M87), 
the orientation of the disk and the detailed shape of the 
perturbations will enormously constrain the model parameters.
Dynamical models are therefore a precious tool to investigate 
on the observed perturbations.

\begin{acknowledgements}
We would particularly like to thank J.R. Roy and P. Martin for making
their H$\alpha$ data available. We thank V. Rubin for making available
her data prior to publication, J. Braine for fruitful discussions, and
the referee for useful advices.
We thank also the VLA staff for their kind support
during the observations.\\
BV is supported by a TMR Programme of the European Community
(Marie Curie Research Training Grant). 
\end{acknowledgements}

\end{document}